\documentclass{article}

\usepackage{arxiv}

\usepackage[utf8]{inputenc} 
\usepackage[T1]{fontenc}    
\usepackage{hyperref} \hypersetup{hidelinks}    
\usepackage{url}            
\usepackage{booktabs}       
\usepackage{amsfonts}       
\usepackage{nicefrac}       
\usepackage{microtype}      
\usepackage{lipsum}		
\usepackage{graphicx}
\usepackage[numbers,sort&compress]{natbib}
\usepackage{doi}

\title{Does Research Software Agree with Itself? A Multi-Surface Consistency Audit of Software Citation Metadata}

\author{ \href{https://orcid.org/0009-0009-6309-380X}{\includegraphics[scale=0.06]{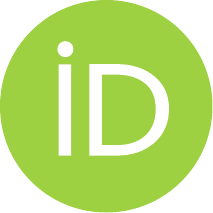}\hspace{1mm}Pengyin Shan} \\
	National Center for Supercomputing Applications\\
	University of Illinois Urbana-Champaign\\
	Urbana, IL 61820 \\
	\texttt{pengyins@illinois.edu}
}

\renewcommand{\shorttitle}{A Multi-Surface Consistency Audit of Software Citation Metadata}

\hypersetup{
pdftitle={A Multi-Surface Consistency Audit of Software Citation Metadata},
pdfsubject={computer science, research software engineering, software engineering},
pdfauthor={Pengyin Shan},
pdfkeywords={software citation, metadata consistency, research software, CITATION.cff, provenance, scholarly infrastructure},
}

\begin{document}
\maketitle

\begin{abstract}
Research software projects describe themselves in many places at once: citation files in the repository, archive deposits, DOI registry records, package registries, and README text. We treat the software as the underlying object and these machine-readable self-descriptions as its surfaces: the points where people and automated systems read what the project declares about the software. Citation guidance, indexing services, and automated agents may read a different subset of these surfaces, so disagreement between them can silently fragment credit and provenance. This paper asks a simple question that has not been measured directly: when a project's own metadata surfaces are compared with each other, how often do they agree? We audited 117 open-source research software projects, comprising an 87-project high-performance computing and quantum computing corpus and a 30-project registered baseline drawn from the JOSS and pyOpenSci accepted-package lists, across up to seven machine-readable surfaces per project. Using a four-level verdict rubric across six metadata fields, with 98.5\% hand-verified verdict precision on a 338-row stratified sample, we found that 52 of the 62 projects exposing at least two comparable surfaces (83.9\%) contain at least one core-field conflict, a result that is insensitive to the fuzzy-matching threshold. Half of hand-adjudicated cross-surface conflicts trace to a single mechanism: surfaces describing the software's paper rather than the software itself. Among projects whose CITATION.cff includes a preferred citation, 28 of 32 route citations to a record that disagrees with the software's own metadata. The author lists and titles disagree the most, and the registry surfaces are the least aligned. We release the audit pipeline as an importable library, the corpus, the registered sampling protocol, all raw snapshots, and the complete verification log.
\end{abstract}


\section{Introduction}
The software citation infrastructure has largely solved the problem of \textit{where} a project can declare its citation metadata. CITATION.cff  \cite{druskat_2021_5171937}, codemeta.json \cite{jones2017codemeta}, archived deposits with DOIs \cite{datacite2024schema}, package registry metadata, and README citation sections are all established, machine-readable, and widely promoted. What this infrastructure does not guarantee is that these surfaces say the same thing. Each could be written at a different time, by a different process, for a different consumer: a CITATION.cff is edited by software developers or maintainers, a Zenodo record may be minted automatically from a GitHub release, a PyPI record is generated at the packaging time, and a README BibTeX block is often pasted directly from a paper's landing page and never updated.

Consumers of citation metadata, whether human authors following a "how to cite" section, automated indexers resolving a DOI, or Large Language Models (LLMs) assembling citations, each read one surface and trust it. If the surfaces disagree, which record a citation carries depends on which surface the consumer happened to read. The consequences are familiar from software citation research: split citation counts, misattributed credit, version ambiguity, and paper-software conflation \cite{smith2016principles, howison2016software, katz2018software}. What has been missing is a direct measurement of how often a project's own self-descriptions disagree, across surfaces, at scale, with verified instrument precision.

This paper contributes that measurement. We framed the problem as a \textit{supply-side} audit: we frame what projects declare across multiple machine-readable surfaces they expose, and score pairwise consistency. Our contributions:
\begin{itemize}
	\item \textbf{An audit instrument.} An open pipeline (harvest, normalize, compare, report) covering seven surfaces: CITATION.cff (root and preferred-citation), codemeta.json, .zenodo.json, the resolved DOI record, PyPI and npm registry records, and README citation content. The comparison rubric has four verdict levels over six fields, with all thresholds documented and released.
	\item \textbf{A measured corpus with a registered baseline.} 87 HPC and quantum computing projects from a prior, DOI-anchored corpus \cite{shan2026channel}, plus a 30-project baseline drawn from the accepted-package lists of JOSS and pyOpenSci under a sampling protocol fixed and timestamped before candidate acceptance.
	\item \textbf{Verified findings.} 83.9\% of projects with at least two comparable surfaces disagree with themselves on at least one core field, with verdict precision hand-verified at 98.5\% over a stratified 338-row sample and results insensitive to the matching threshold.
    \item \textbf{A mechanism taxonomy.} Half of adjudicated cross-surface conflicts trace to paper-versus-software conflation, and the remainder cluster into archive staleness, registry ghost metadata, and identity ambiguities, each illustrated with named, snapshot-dated examples.
\end{itemize}

\section{Related Work}
\label{sec:related_work}

The FORCE11 software citation principles established software as a citable research product and specified what a citation must identify: the artifact actually used, through its authors, title, version, date, and a unique persistent identifier \cite{smith2016principles}. The FAIR principles for research software carry the same position into metadata practice, tying findability and reuse to rich, machine-readable description of the software itself \cite{barker2022fair}. Reviewing the implementation two years after the principles, Katz and Chue Hong reported that the infrastructure had arrived, with over 50,000 DOIs issued for software at the time, while challenges remained in developer practice and in publisher systems retaining the required metadata \cite{katz2018software}. Five of the six fields our rubric compares (title, authors, version, year, and DOI) are the fields this line of work designates as citation-critical; the sixth, license, is carried by every machine-readable surface we audit. Our question is not whether projects declare these fields but whether the surfaces declaring them agree.

Declaration infrastructure followed the principles. The Citation File Format gives maintainers a plain-text citation file in the repository root \cite{druskat_2021_5171937}, and GitHub renders that file as the "cite this repository" widget on the repository landing page, so a single YAML file now reaches every visitor. CodeMeta provides a JSON-LD exchange schema with crosswalks between registry and archive vocabularies \cite{jones2017codemeta}, and archive DOI records are registered against the DataCite metadata schema \cite{datacite2024schema}. Adoption of this infrastructure is measured at registry scale: El Hounsri and Garijo analyzed research software metadata adoption across the software registries of five European Open Science clusters and a multi-disciplinary registry and found recommended practice unevenly followed \cite{11025739}. Adoption studies of this kind have established whether a surface is present and well-formed, but none has asked whether the surfaces a project has adopted describe the same object in the same way, which is the measurement this paper adds.

Where metadata quality is studied, it is studied one surface at a time. On the paper side, Howison and Bullard examined how software is mentioned in the biology literature and found mentions that were informal and incomplete, lacking the version and identifier information needed to find the artifact used \cite{howison2016software}. On the registry side, Bommarito and Bommarito characterized package metadata across 178,592 PyPI packages \cite{bommarito2019pypi}, and Gao et al. measured that existing tools can recover source-repository information for at most 70.5\% of PyPI releases because release metadata omits the repository or names the wrong one, and built PyRadar to retrieve and validate those links \cite{gao2024pyradar}. The newest large-scale datasets aggregate in the same direction: OpenDORS compiles license, version, and language metadata for over 120,000 openly referenced research software repositories, one record per project \cite{druskat2025opendors}. Each of these efforts measures the presence or quality of a single surface across many projects, or merges a project's surfaces into one record. To our knowledge, no prior study compares a project's own self-description surfaces against each other, so the rate at which research software disagrees with itself has not been measured. That distinction also separates this audit from adoption analyses such as \cite{11025739} as counting surfaces and cross-checking them are different measurements, and only the first exists in the literature.

A separate line of work documents the paper standing in for the software. The Journal of Open Source Software aims to create a peer-reviewed, citable paper, so that software work can be credited through existing publication infrastructure \cite{katz2018publish}; which citation carries is left to authors and citing practice. Corpus studies show how the proxy pattern emerges in that practice. The Softcite dataset, built from expert annotation of 4,971 biomedical and economics publications, found that software is mentioned informally far more often than it is formally cited \cite{du2021softcite}, and a follow-up study of the CORD-19 corpus found that even where projects request a specific citation, identifying the canonical record among a project's several candidate sources is difficult \cite{du2022understanding}. These studies measure conflation where it lands in the citing literature. Our mechanism analysis in Section \ref{sec:A mechanism taxonomy of verified conflicts} measures it one step earlier: in the machine-readable metadata projects themselves publish, where half of verified cross-surface conflicts trace to a surface describing the paper rather than the software.

Supply-chain security frameworks audit adjacent layers of the same projects. OpenSSF Scorecard scores repository and maintenance practices automatically \cite{zahan2023scorecard}; SLSA defines graded levels of build provenance, and since the specification is community-maintained without a peer-reviewed primary description, we reference the qualitative study of its deployment challenges by Tamanna et al. \cite{tamanna2024slsa}; Sigstore provides signing infrastructure for released artifacts \cite{newman2022sigstore}. These frameworks address whether code and artifacts are what they claim to be. None checks whether a project's self-descriptions agree with each other, so the audit reported here is complementary to them and overlaps none of their controls.

\section{Methods}
\label{sec:methods}
\subsection{Corpus}
\label{sec:corpus}
The corpus contains 117 projects in three strata. The \textbf{supercomputing stratum} (labeled \verb|sc26| in the released data files; 87 projects; 44 HPC, 43 quantum computing) is exported verbatim from the tagged v1.0.0 release of our prior provenance-audit corpus \cite{shan2026channel}, where the derivation from the tagged release keeps the corpus identity claim checkable. The \textbf{baseline strata} (15 JOSS, 15 pyOpenSci) exist to test whether findings from infrastructure-heavy HPC/Quantum software generalize to community-reviewed research software, and were drawn as follows.

Sampling frames were retrieved on 2026-08-12 from the JOSS published-papers API (3,662 records after retrieval) and the pyOpenSci packages registry (70 records). Corresponding software repositories were required to be GitHub-hosted, since all six eligibility surfaces are probed through the GitHub API. Non-GitHub candidates were excluded at sampling under rule R5 (JOSS: 185, pyOpenSci: 4, hosts reported in the exclusion log). Four further pre-registered rules applied in order: R1 corpus overlap with the 87 (7 exclusions), R2 dual-listing resolved to the pyOpenSci stratum (32), R3 JOSS version papers deduplicated to the most recent (22), R4 missing repository URL (0). Domain bins were assigned from a committed keyword map (first match wins; disciplinary keywords precede generic ones). Candidates with no keyword matched were retained in an explicit "unmapped" bin that participates in the stratified draw rather than being excluded, so the draw does not depend on the coverage of a hand-written map. The registered unmapped share was 33\% (JOSS) and 2\% (pyOpenSci). One design correction was made before the protocol was registered: an earlier rule that refused to finalize while unmapped candidates appeared in the draw window was unsatisfiable under round-robin interleaving, since the unmapped bin is guaranteed window slots by construction, and it was replaced by unmapped-bin participation plus a 50\% frame-level ceiling.

A seeded, domain-stratified draw (seed 202608121338, round-robin over domain bins, oversample 30 per stratum) produced ranked candidate windows. The sampling protocol, ranked candidate lists, and exclusion log were committed (baseline-sampler v0.1.0) and deposited under restricted access with a DOI before any eligibility probing (10.5281/zenodo.21909720; the deposit timestamp precedes the acceptance log timestamp). Eligibility required at least 2 of six machine-readable surfaces, probed per candidate: CITATION.cff, codemeta.json, .zenodo.json in the repository root, a README citation surface (citation section or BibTeX block; the registered rule), a discoverable DOI (from CFF or the README citation section, not resolved at probe time) and a verified registry listing (a PyPI or npm package matching the repository name and linking back to the repository). Acceptance proceeded strictly in rank order to 15 per stratum: JOSS accepted ranks 2-18 (3 skips for insufficient surfaces), pyOpenSci ranks 1-19 (4 skips). One drawn candidate's declared repository did not resolve at probe time and was never reached by acceptance. 15 of the 30 pyOpenSci window candidates are dual-listed with JOSS and therefore carry a JOSS paper DOI, so the DOI-by-construction confound noted below extends partially into that stratum.

Hand-curated concept DOIs were filled for the supercomputing stratum only (20 projects), drawn from the prior corpus's distribution-metadata records; both baseline strata rely entirely on automatic discovery from project surfaces, keeping the baseline availability numbers free of curation effects. Registry package names were detected uniformly for all 117 projects by name-part lookup verified by repository back-link and were hand-reviewed. Two detections were removed as packages not controlled by the project.

\subsection{Surfaces}
\label{sec:surfaces}

For each project we harvest up to seven surfaces: \verb|cff| (CITATION.cff root), \verb|cff_preferred| (its preferred-citation block, extracted as a pseudo-surface so intra-file disagreement is measurable), \verb|codemeta| (codemeta.json), \verb|zenodo_json| (.zenodo.json), \verb|doi_record| (the DOI's registered metadata, resolved via DataCite \cite{datacite2024schema} with Crossref fallback; DOI priority order: corpus-curated concept DOI, then CFF-declared, then README citation section, with provenance recorded), \verb|pypi| and \verb|npm| (registry records), and \verb|readme| (a heuristic parse of the README's citation section, BibTeX block, and DOIs). GitHub's "cite this repository" widget is generated from CITATION.cff and is covered by the \verb|cff| surface. The README surface is defined as a README file in the repository root; GitHub additionally renders docs/ and .github/ READMEs on the landing page, which this audit does not credit (one corpus project is affected). Every API response is snapshotted with a retrieval timestamp; all snapshots were retrieved on 2026-08-14 and are released in the data deposit.

\subsection{Normalization}
\label{sec:normalization}

Each surface is normalized into one record shape: title, authors (family, given, ORCID), version, year, DOI, license. Extraction rules are released with the pipeline, where three deserve note: BibTeX in READMEs is extracted with brace-matching rather than regular expressions, after live testing showed regex extraction silently dropping fields across common forms; DOI strings are sanitized (resolver-URL prefixes, badge-image suffixes, trailing punctuation) and validated against the DOI pattern before resolution; DataCite creator records frequently carry the full name in familyName with no givenName so the normalizer uses record-supplied name parts as-is and falls back to name-splitting only when the record supplies neither part.

\subsection{Comparison rubric}
\label{sec:Comparison rubric}

For every pair of surfaces present, each of the six fields receives one of four verdicts. \textbf{exact}: normalized values identical. \textbf{minor}: clearly the same referent with superficial variation (title token-sort fuzzy >= 90; the same author set reordered or fuzzy-matched; year differing by exactly 1; version differing only by a v prefix). \textbf{conflict}: both present and materially different. \textbf{missing}: at least one side absent. Zenodo concept-versus-version DOI pairs are flagged for hand adjudication rather than auto-resolved. All thresholds live in one released module.

Headline results are insensitive to the fuzzy-match threshold: at 85 and 95 the share of projects whose surfaces contain at least one core-field conflict is unchanged (83.9\%, with identical per-stratum rates), while field-level agreement shifts modestly (authors 0.459-0.511 around the reported 0.481; title 0.436-0.442; median per-project pairwise agreement 0.50-0.55). Threshold-sensitivity outputs are released alongside the results.

\subsection{Verification protocol and instrument disclosure}
\label{sec:Verification protocol and instrument disclosure}

We hand-verified the instrument's verdicts against the harvested snapshots rather than the live web, since precision measures whether the pipeline judged correctly given the evidence it retrieved. The verification set comprised every conflict verdict, every minor verdict, every intra-file conflict between CFF root and preferred-citation metadata, and a seeded random sample of 30 exact verdicts. For each judgment, we re-derived the verdict by hand from the recorded value pair; identical value pairs recurring across surface pairs within a project were adjudicated once and the judgment propagated to each covered comparison row. Raw snapshots were consulted where a recorded value appeared anomalous. Rows flagged as possible Zenodo concept-versus-version identifier pairs were additionally adjudicated by resolving both identifiers.

Verified precision was 98.5\% over 338 comparison rows (275 deduplicated judgments): conflict 276/281 (98.2\%), minor 26/26, exact 31/31. The five disagreeing rows classify into two mechanisms, both documented heuristic limits: threshold boundary (2 rows; an author-entity name variant scoring just below the fuzzy threshold) and parse limitations (3 rows; an organizational author split at an embedded "and", and a LaTeX-escaped accented character blocking one name match). The complete verification log is released.

Instrument defects found and corrected before the reported measurements, in the honest-reporting style this protocol requires: two normalization defect classes were identified in pre-run live testing (YAML date serialization in CITATION.cff snapshots; .zenodo.json license objects producing spurious license conflicts); three DOI-string hygiene classes were identified in first-run review (resolver-URL prefixes, badge-image suffixes, non-DOI identifier values); and hand verification identified one systematic defect (DataCite creator records carrying full names in familyName caused token duplication and depressed author matching), which was corrected and the affected comparisons re-scored and re-verified. The instrument's stage history (v0.2.0 pre-run; v0.2.1 first-run corrections; v0.2.2 verification-corrected final) is documented above and in the repository README; the public repository history was flattened for release, and v0.2.2 produced every number reported here.

\section{Results}
\label{sec:results}

\begin{table}[t]
\centering
\caption{Availability of self-description surfaces across the corpus
(n = 117; all projects resolvable at snapshot time).}
\label{tab:availability}
\begin{tabular}{lr}
\toprule
Surface & Availability \\
\midrule
README (root-level)\textsuperscript{a}   & 116/117 (99\%) \\
Stable releases                          & 111/117 (95\%) \\
CITATION.cff\textsuperscript{b}          & 48/117 (41\%)  \\
Registry listing (PyPI or npm)\textsuperscript{c} & 46/117 (39\%) \\
Archival DOI (doi\_record)               & 20/117 (17\%)  \\
SECURITY.md                              & 16/117 (14\%)  \\
.zenodo.json\textsuperscript{d}          & 10/117 (9\%)   \\
codemeta.json                            & 6/117 (5\%)    \\
\bottomrule
\end{tabular}
\vspace{2pt}
\begin{minipage}{\linewidth}
\footnotesize
\textsuperscript{a}\,One project (preCICE) keeps its README under docs/;
under the root-only rule it is counted absent (see Section 3.2).
\textsuperscript{b}\,By stratum: 26/87 supercomputing, 10/15 JOSS, 12/15 pyOpenSci.
\textsuperscript{c}\,One project is npm-listed; it is also PyPI-listed.
\textsuperscript{d}\,By stratum: 3/87 supercomputing, 1/15 JOSS, 6/15 pyOpenSci.
\end{minipage}
\end{table}

\subsection{Surface availability}
\label{sec:Surface availability}

All 117 repositories resolved at snapshot time. Table~\ref{tab:availability} reports surface availability. The human-facing substrate is near-universal: README in 116 of 117 projects and stable releases in 95\%. The machine-readable citation surfaces are minority practice, led by CITATION.cff at 41\% and thinning to codemeta.json at 5\%. 62 of 117 projects (53\%) expose at least two comparable surfaces; the consistency results below concern those 62. That only half the corpus can be cross-checked at all is itself a finding: for the other half, whatever the single surface says is uncheckable by construction.

\subsection{Pairwise consistency}
\label{sec:Pairwise consistency}

\begin{figure}
	\centering
	\includegraphics [width=\textwidth]{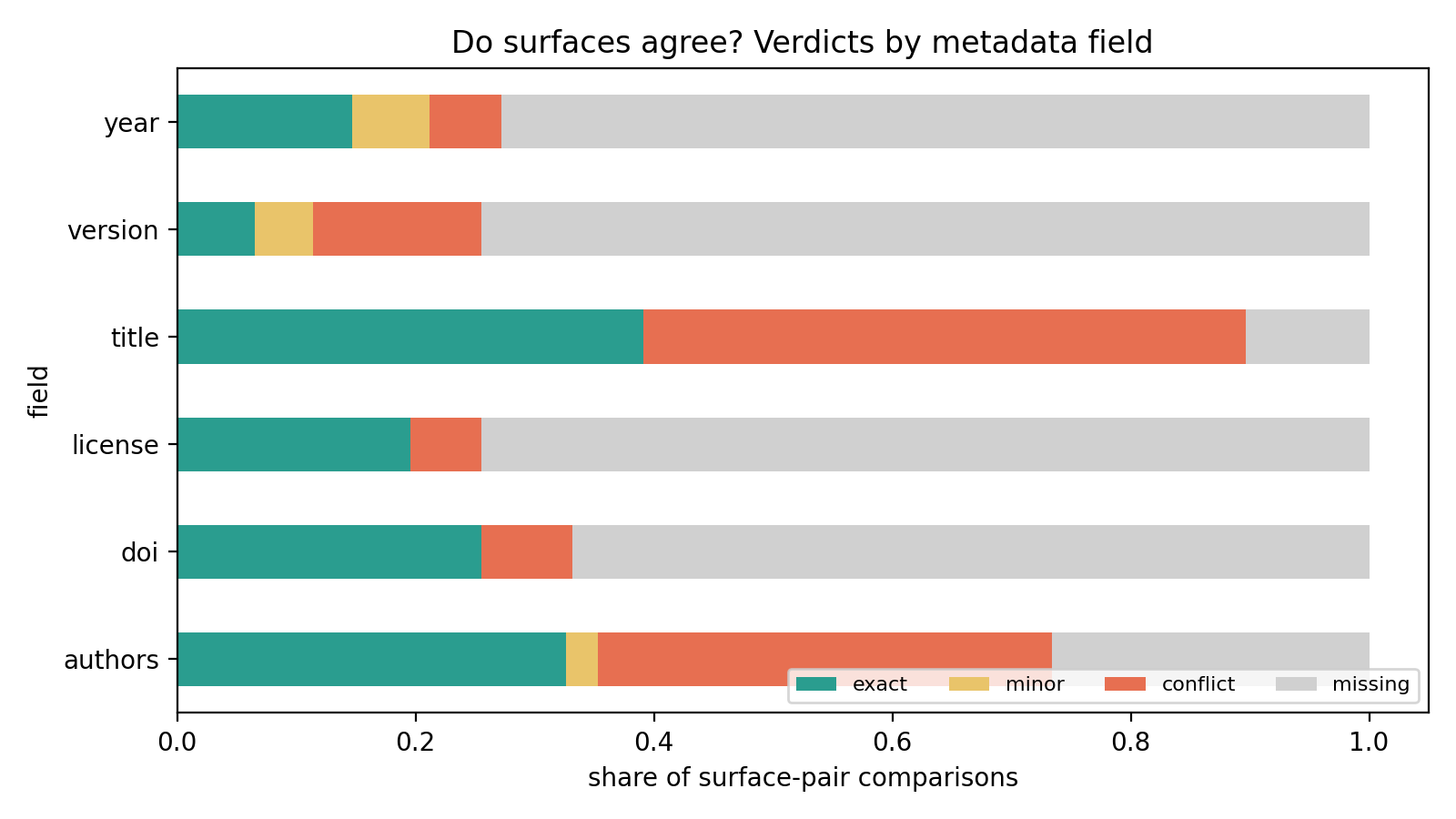}
	\caption{Verdict composition per field across all 1,104 surface-pair field comparisons. Gray indicates the field was not comparable because at least one surface omits it; Table 2 reports agreement among the comparable remainder.}
	\label{fig:fig1}
\end{figure}

\begin{figure}
	\centering
	\includegraphics [width=\textwidth]{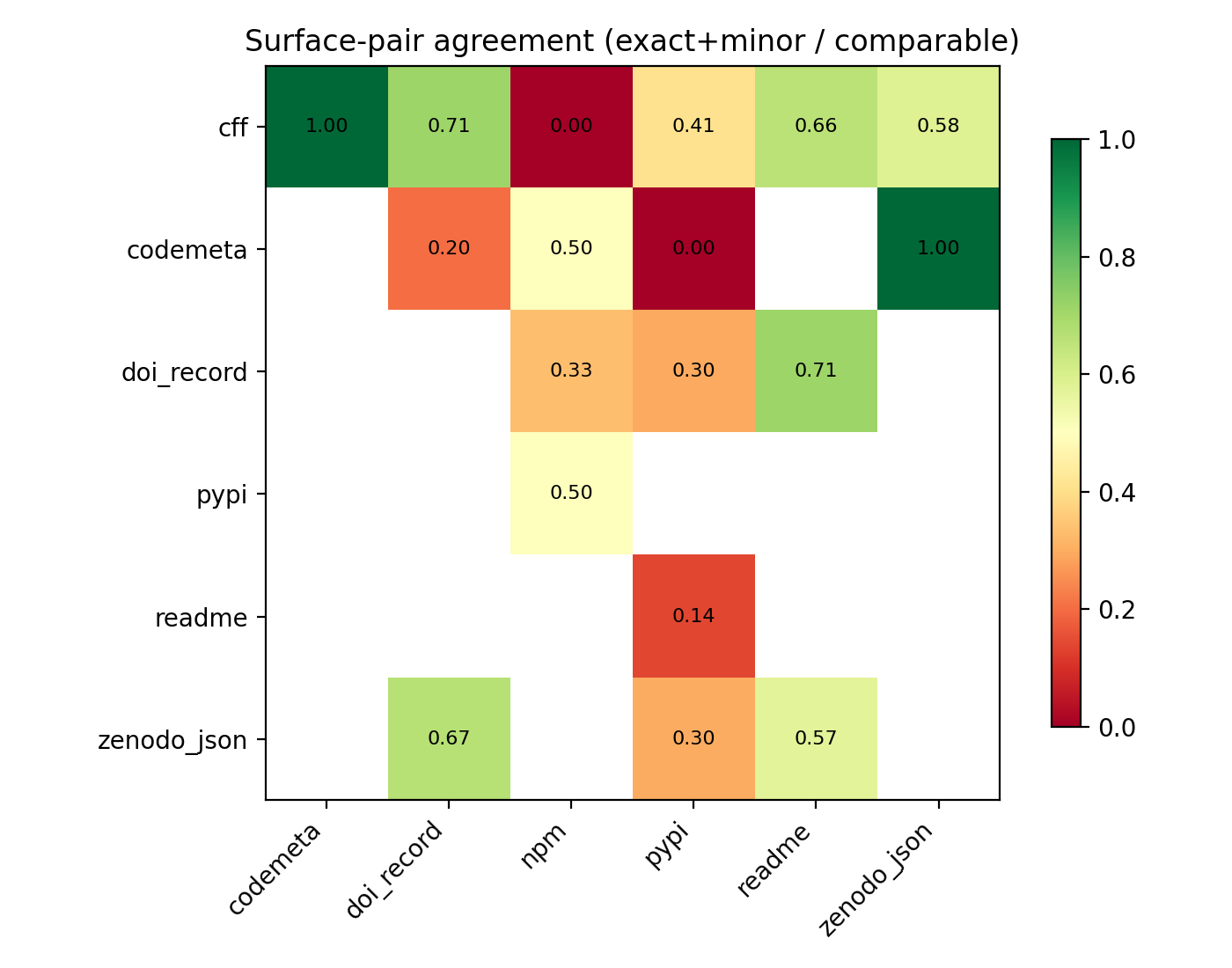}
	\caption{Pairwise agreement (exact plus minor over comparable) by surface pair. Cell denominators vary widely; extreme values, including 0.00 and 1.00, can rest on few comparable pairs, and per-cell counts are in the released results data; the corpus contains a single npm-listed project.}
	\label{fig:fig2}
\end{figure}

\begin{table}[t]
\centering
\caption{Pairwise agreement by field across comparable surface pairs,
with and without registry-involved pairs. Registry exclusion is shown
only for the two fields where registry pairs change the rate materially.}
\label{tab:fields}
\begin{tabular}{lrrrr}
\toprule
 & \multicolumn{2}{c}{All comparable pairs} & \multicolumn{2}{c}{Excluding registry pairs} \\
\cmidrule(lr){2-3}\cmidrule(lr){4-5}
Field   & $n$ & Agreement & $n$ & Agreement \\
\midrule
Year    & 50  & .780 &     &      \\
DOI     & 61  & .770 &     &      \\
License & 47  & .766 &     &      \\
Authors & 135 & .481 & 98  & .612 \\
Version & 47  & .447 &     &      \\
Title   & 165 & .436 & 90  & .644 \\
\bottomrule
\end{tabular}
\begin{minipage}{\linewidth}
\vspace{2pt}\footnotesize
Of 93 title conflicts, 61 involve a registry surface.
\end{minipage}
\end{table}

Across 1,104 field comparisons (505 comparable, 599 with at least one side missing), 52 of 62 projects (83.9\%) contain at least one core-field conflict (title, authors, version, year, or DOI), and median per-project pairwise agreement is 0.50. By field, agreement among comparable pairs is: year 0.78, DOI 0.77, license 0.77, authors 0.48, version 0.45, title 0.44. Figure~\ref{fig:fig1} shows the composition behind these rates: of 1,104 field comparisons, 599 could not be made at all because at least one surface omits the field, so the agreement rates in Table~\ref{tab:fields} describe only the comparable remainder.

The registry surfaces drive much of the title and authors disagreement: 61 of 93 title conflicts involve PyPI or npm, largely the structural gap between a registry's package name and a descriptive title ("mpi4py" versus "MPI for Python"). Excluding registry pairs, title agreement rises from 0.44 to 0.64 and authors from 0.48 to 0.61. Figure~\ref{fig:fig2} locates the disagreement: pairs involving a registry surface concentrate the low agreement values, while the repository-native pairs (cff with codemeta, codemeta with .zenodo.json) agree most. We report both because both are real: a human reader distinguishes a package name from a title, but a metadata consumer joining records across surfaces sees a mismatch.

\begin{table}[t]
\centering
\caption{Consistency by stratum among projects with at least two
comparable surfaces. Agreement is pooled pairwise agreement: exact plus
minor verdicts over comparable pairs within the stratum.}
\label{tab:strata}
\begin{tabular}{lrrrr}
\toprule
Stratum   & Projects & Comparable pairs & Agreement & $\geq$1 core-field conflict \\
\midrule
supercomputing      & 36 & 236 & .559 & 86.1\% \\
JOSS      & 11 & 97  & .649 & 63.6\% \\
pyOpenSci & 15 & 172 & .494 & 93.3\% \\
\midrule
Overall   & 62 & 505 & .554 & 83.9\% \\
\bottomrule
\end{tabular}

\vspace{2pt}
\begin{minipage}{\linewidth}
\footnotesize
The median per-project pairwise agreement across the 62 projects is .500.
\end{minipage}
\end{table}

By stratum, the phenomenon is not an HPC idiosyncrasy: core-field conflict affects 86.1\% of supercomputing projects (31/36 with >= 2 surfaces), 63.6\% of JOSS projects (7/11), and 93.3\% of pyOpenSci projects (14/15). Two cautions attach to the baseline comparison: the JOSS stratum's DOI field is partially DOI-by-construction (every JOSS paper carries a DOI, and 15 of 30 pyOpenSci window candidates are dual-listed), and per-stratum denominators are small (11-36 projects), so we report the direction (community-reviewed packages are not cleaner; pyOpenSci projects expose more surfaces and correspondingly more disagreement) without significance claims. One tautology is excluded from interpretation: when the resolved DOI was discovered from a project's own CFF or README, the DOI-field agreement between the discovering surface and the DOI record is expected; informative DOI comparisons are those between non-source surfaces, including the four adjudicated concept-versus-version pairs.

\subsection{The project's own citation file disagrees with itself}
\label{sec:The project's own citation file disagrees with itself}

32 projects' CITATION.cff files include a preferred-citation block. In 28 of 32 (87.5\%), following the preferred-citation yields a record that conflicts with the software's own metadata on at least one field (56 conflicting field comparisons; 39 exact, 1 minor). This is partly the CFF specification working as designed, since preferred-citation exists precisely to route citation to a different object, usually a paper. The measured fact is about the consequence: for seven out of eight such projects, the record a compliant citation tool emits and the record the software's other surfaces describe are different objects, and which one a consumer lands on depends on tooling, not on the project's intent.

\subsection{A mechanism taxonomy of verified conflicts}
\label{sec:A mechanism taxonomy of verified conflicts}

Hand adjudication of every conflict produced a mechanism classification; the four recurring mechanisms, with named, snapshot-dated examples (retrieved 2026-08-14):

\paragraph{Paper-versus-software conflation (86 of 171 cross-surface conflict judgments, 50\%).}
Surfaces describe the software's paper rather than the software. Two sub-patterns. Divergent: dace's CITATION.cff root titles the software ("DaCe - Data Centric Parallel Programming") while its README BibTeX carries the conference paper's title and its five authors, against 28 in the CFF. \textit{Consistent-but-paper-anchored:} thread-pool's CFF root itself declares the SoftwareX paper's DOI and title, so its surfaces agree with each other while all describing the paper; consistency is not correctness of referent.

\paragraph{Archive staleness.}
DOI'd archive records lag the distributed release: felupe's Zenodo record says v9.4.0 while PyPI serves 10.1.0; cuda-quantum shows the same mechanism at minimum distance (0.15.0 versus 0.15.1). The rubric counts each non-equal pair once regardless of distance.

\paragraph{Registry ghost metadata.}
Registry records exist that the project does not control or no longer describes: our review removed two back-link-verified packages (a community npm build and a third-party PyPI wrapper) as not project-controlled; registry author fields are frequently entity strings or empty, and no registry surface in the corpus carries an ORCID (0/25 sampled authors on PyPI).

\paragraph{Identity ambiguity.}
Entity and person names collide across surfaces: cirq's surfaces disagree only on "Cirq Developers" versus "The Cirq Developers"; mfem's CFF root credits "MFEM Team" while its archive record credits two named individuals and its preferred citation seventeen.

By design, the by-design routing of Section \ref{sec:The project's own citation file disagrees with itself} (mfem-style preferred-citation) is tabulated separately from cross-surface conflation, since the former is specification-compliant behavior with a fragmenting consequence and the latter is unintended divergence.

\subsection{ORCID coverage}
\label{sec:ORCID coverage}
Where surfaces carry authors, ORCID coverage varies from 62\% of authors in .zenodo.json and 54\% in CFF to 39\% in resolved DOI records and 0\% in registry records and README BibTeX, so the surfaces most likely to be read by packaging tooling carry the least persistent-identifier information.

\section{Discussion}
\label{sec:Discussion}
Three implications follow from the supply side alone. First, single-surface consumers inherit whichever record their surface carries; at the measured disagreement rates, two compliant consumers reading different surfaces of the same project will disagree about its citation roughly half the time at the field level. Second, the paper-versus-software mechanism dominating verified conflicts (50\%) means the fragmentation is not primarily sloppiness: it is the unresolved question of what the citable object is, propagating into machine-readable metadata, consistent with the credit incentives documented in software citation research \cite{smith2016principles, katz2018software}, and it will not be fixed by adoption campaigns for any single format. Third, agreement infrastructure is absent: no existing supply-chain or metadata framework checks cross-surface consistency (\cite{zahan2023scorecard, tamanna2024slsa, newman2022sigstore} audit build and release integrity, not self-description coherence), and the near-universal surfaces (README, releases) are precisely the ones with the least structure.

The audit engine is released as an importable library whose per-project primitive (harvest, normalize, compare, report for one repository) is designed for downstream reuse in release-time integrity checks and citation and provenance checker tools; the corpus-scale study reported here is one instantiation. The demand side, how automated consumers weight these surfaces when they disagree, is measurable with the same instrument as ground truth and is deliberately out of scope here.

\section{Limitations}
\label{sec:Limitations}

The corpus is GitHub-hosted by criterion and skewed toward HPC/QC by design, with small baseline strata. README parsing is heuristic; verified precision (98.5\%) bounds but does not eliminate instrument error, and the released log classifies every known disagreement. The registry surface covers PyPI and npm only, so R and Julia packages face a five-surface ceiling, and their registry practice is unmeasured. Registry name-versus-title conflicts are structural and reported both included and excluded. The eligibility filter conditions the baseline on having >= 2 surfaces, so baseline agreement rates describe metadata-active projects, not the registries at large. Snapshots date to a single retrieval window; projects change. Concept-versus-version DOI adjudication covered flagged Zenodo pairs only. Verdict counts weight all conflicts equally regardless of severity.

\section{Conclusion}
Measured across their own machine-readable surfaces, five of six research software projects that can be cross-checked disagree with themselves on at least one citation-critical field, and the single largest verified mechanism is the paper standing in for the software. The instrument, corpus, registered protocol, snapshots, and verification log are released so the measurement can be repeated, extended to other ecosystems, and used as ground truth for the demand-side question of what metadata consumers do when the surfaces disagree. Extending the corpus beyond GitHub-hosted projects and additional registries, and re-measuring the same projects over time and assessing mitigation in which a single authoritative record generates the other surfaces, is planned future work on the same instrument.

\section{Data and Code Availability}
Pipeline: \href{https://github.com/pengyin-shan/rda-audit-pipeline}{https://github.com/pengyin-shan/rda-audit-pipeline} (Apache-2.0). Measurements were produced with v0.2.2; the archived release v0.2.3 adds only license, citation, and README files, with measurement code identical (10.5281/zenodo.21969695). Corpus, raw snapshots (retrieval-timestamped), results, verification log, and sensitivity outputs: \textit{10.5281/zenodo.21969769}, CC-BY-4.0. Sampling protocol, ranked candidate lists, and exclusion log: \textit{10.5281/zenodo.21909720} (restricted until publication; opened on posting); the same files are public in the sampler repository. Sampler: baseline-sampler v0.1.0 (registered draw commit 64b979e), \href{https://github.com/pengyin-shan/baseline-sampler}{https://github.com/pengyin-shan/baseline-sampler}. Prior corpus: \textit{10.5281/zenodo.21443211}.

\section{Acknowledgments}
The author thanks Daniel S. Katz for comments on a draft. Portions of the manuscript text were drafted with writing assistance from Claude (Anthropic); the author reviewed and verified all content, and all data collection, measurement, and verification were performed by the author. The corpus derives from work conducted during the author's 2025-2026 Trusted CI Fellowship with Trusted CI, the NSF Cybersecurity Center of Excellence; the fellowship provided no financial support for this work. The author thanks the National Center for Supercomputing Applications (NCSA), University of Illinois Urbana-Champaign, for a supportive professional environment; this work was conducted in the author's individual capacity and received no dedicated funding. The author declares no competing interests.

\bibliographystyle{unsrtnat}
\bibliography{references}

\end{document}